\documentclass{appolb}
\usepackage{epsfig}
\begin{document}
\title{Exotic Mesons, Theory and Experiment
\thanks{Presented at MESON2000 (Cracow, 19-23 May 2000)}
}
\author{T.Barnes
\address{Physics Division,  
Oak Ridge National Laboratory\\
Oak Ridge, TN 37831-6373, USA  \\
Department of Physics and Astronomy,
University of Tennessee\\
Knoxville, TN 37996-1501, USA \\
Institut f\"ur Theoretische Kernphysik
der Universit\"at Bonn \\
Nu\ss allee 14-16,
D-53115 Bonn, Germany\\
Institut f\"ur Kernphysik, 
Forschungszentrum J\"ulich\\
D-52425 J\"ulich, Germany\\ }     
}
\maketitle
\begin{abstract}
In this talk I summarize the status of exotic mesons, including
both theoretical expectations and experimental candidates.
The current experimental candidates are  ``spin-parity exotics'';
since these are most often considered possible
hybrid mesons, the
theoretical discussion will be mainly concerned with hybrids.
The exotic meson candidates discussed are the surprisingly
light $\pi_1(1400)$ and $\pi_1(1600)$.
\end{abstract}
  
\section{Theoretical Expectations}
\subsection{Exotics Defined}

An ``exotic meson'' has J$^{PC}$ or flavor quantum numbers forbidden
to the 
$|q\bar q\rangle $ states of the
nonrelativistic quark model.

The current experimental candidates are 
``spin-parity exotics'', which have J$^{PC}$ forbidden to
$q\bar q$ mesons. In principle one might also find flavor exotics
in a multiquark sector, for
example in I=2, but no such 
(widely accepted) experimental candidates are known at present \cite{GN}.

As a {\it caveat} we emphasise that every meson is 
a linear superposition of all allowed
basis states, spanning $|q^2\bar q^2\rangle$, $|q\bar q g  \rangle$, 
$|gg  \rangle, \dots $ (where not strictly forbidden), 
with amplitudes that
are determined by QCD interactions. For convenience we usually classify
resonances as ``quarkonia'', ``hybrids'', ``glueballs'' and so forth,
and are implicitly assuming that one type of basis state dominates the state
expansion of each resonance. Of course this may not be the case in general,
and the amount of ``configuration mixing'' is an open and rather 
controversial topic in hadron physics.
Exotics are special in that the $|q\bar q\rangle$ component must be zero,
due to the quantum numbers of the state.

\subsection{What became of multiquarks?}

Multiquark systems such as $q^2\bar q^2$ were once expected to contribute a
rich spectrum of resonances to the meson spectrum, and in the 1970s there
were many detailed calculations of the spectrum of multiquark 
resonances
in various models. Now one hears little
about this subject. What became of multiquarks?

The answer is that they ``fell apart''. Even in the early work on 
$q^2\bar q^2$ multiquarks \cite{Jaffe1}
it was realized that their decay couplings would be very different from
conventional $q\bar q$ mesons; the latter decay 
mainly through the production of 
a second $q\bar q$ pair, whereas the $q^2\bar q^2$ system
can simply be rearranged
into 
a state of two $(q\bar q) +  (q\bar q)$ mesons.
If the expected energy of a continuously deformed
$q^2\bar q^2 \to (q\bar q)  (q\bar q)$ 
system
is monotonically decreasing, one would not expect to find
a  $q^2\bar q^2$  resonance. This was the situation found 
variationally in the
scalar sector by Weinstein and Isgur for most light quark masses 
\cite{WI}. 

Life can be more complicated, and Weinstein
and Isgur also found that weakly bound deuteronlike 
${\rm K}\bar {\rm K}$ states existed
in their model. 
Presumably many more such 
weakly bound 
quasinuclear states exist, both in meson-meson and meson-baryon sectors.
This subject of multi{\it hadron} systems is at least as rich as the 
table of nuclear levels.

One often hears that the $q^6$ system may have a bound state in
the $u^2d^2s^2$ I=0, J=0 flavor sector, known as the ``H dibaryon''
\cite{Jaffe2}. 
Caution is appropriate here. Some experiments that are 
nominally searching for the H dibaryon have ``widened their net'' to
include
states very close to 
$\Lambda\Lambda$ threshold; if one is found, it
would more likely be a weakly bound $\Lambda\Lambda$ hypernucleus. 
It is important not to equate these two ideas.
A $\Lambda\Lambda$ hypernucleus 
would certainly be a very interesting discovery, 
especially in its implications for
models of the intermediate ranged baryon-baryon attraction, but it
is {\it not} the H dibaryon envisaged in bag model calculations.
The H dibaryon calculations 
assumed an SU(3) flavor-singlet $u^2d^2s^2$ system,
and $\Lambda\Lambda$ is a quite different flavor state.
Quark model
calculations actually find the  $\Lambda\Lambda$ interaction
to be repulsive, rather like the NN core. 

It does appear likely that real multiquark $Q^2\bar q^2$ 
clusters will exist given sufficiently
heavy quarks ($Q=c$, or perhaps only $b$)
\cite{Richard}, but these are unfortunately not easily accessible
to experiment.

\subsection{Hybrid mesons}

A hybrid meson is usually ``defined'' as a resonance whose
dominant valence component is $|q + \bar q + excited \ glue \rangle$.
This deliberately vague definition covers our present ignorance over how
one can most accurately describe gluonic excitation. Possibilities include
models with explicit transverse gluon quanta, such as the bag model, as well
as an excitation of the flux tube that one sees in LGT
simulations. Fortunately for experimenters, many of the model calculations 
reach rather similar predictions for the properties of these states.

One general conclusion is that, unlike $q\bar q$, {\it all} 
J$^{PC}$ can be constructed from hybrid basis states. This conclusion 
is seen most rigorously in the list of gauge-invariant local 
operators that one
may construct from a product of $\bar \psi, \psi$ and $F^a_{\mu\nu}$, since
one may couple to physical states by operating on the vacuum $|0\rangle$
with such an ``interpolating field''.
This list of $\bar \psi \otimes  \psi \otimes F$ operators covers
all J$^{PC}$, including the so-called exotic combinations
\begin{displaymath}
{\rm J}^{PC}\bigg|_{exotic} =
0^{--}, 
0^{+-}, 
1^{-+}, 
2^{+-}, 
3^{-+}, \dots 
\end{displaymath}
that one cannot construct from a  $\bar \psi\otimes \psi$ quarkonium operator.

At present the experimental J$^{PC}$ exotics are usually considered
to be hybrid meson candidates, simply because theorists know of 
no other general class
of J$^{PC}$ exotic resonance, excepting multiquark systems that
purportedly
``fall apart'' into light 
$q\bar q$ mesons. (Possible exceptions which merit future investigation 
are weakly bound quasinuclear states,
which might exist near threshold in S-wave in attractive meson-meson channels.)

In any case, if a  J$^{PC}$ exotic meson is found, we can be certain that
we have discovered something beyond the naive quark model. This is an
extremely important possibility experimentally, and assuming that 
such states are
clearly identified we may hope that the pattern of their spectroscopy will 
eventually make it clear just what has been discovered!

\section{Specific models of hybrids}

\subsection{Introduction}

Much of the work on 
hybrids has made use of very specific models of 
``excited glue''. These models are the bag model, 
the flux tube model, and the rather underexplored constituent gluon model.
Finally, masses and other properties of  J$^{PC}$ exotic hybrids 
may be predicted by QCD sum rules and LGT using
interpolating fields, and these approaches do not make model assumptions 
about the nature of gluonic excitation. We will discuss some of the
more fundamental results of these models, especially as regards masses,
quantum numbers and decay properties.

\subsection{Bag model hybrids}

Many early hybrids studies used the bag model, which assumed
that quarks and gluons could be treated as spherical cavity modes of
Dirac and Maxwell quanta, confined within the cavity by the choice
of color boundary conditions. The ``zeroth-order'' bag model states
were color singlet product basis states of quark, antiquark and
gluon modes, for example
\begin{displaymath}
|q\bar q\rangle \ , \hskip 1cm 
|q\bar qg \rangle \ , \hskip 1cm
|gg \rangle \ , \hskip 1cm
|q^2\bar q^2 \rangle \ , \dots \ .
\end{displaymath}
The quark-gluon and gluon self interactions mixed these basis states,
so that the physical levels were linear combinations of these
``bare'' basis states, just as we anticipated in our introduction.
The distinction between ``conventional $q\bar q$ meson'' and
``nonexotic hybrid'' in the bag model was thus rather vague, and 
was clearest as a theoretical identification as the strength of the
QCD coupling constant was made small. The bag model gave a rather good
description of the light ``conventional $q\bar q$ meson'' spectrum
as $|q\bar q\rangle +  O(\sqrt{\alpha_s}) |q\bar qg \rangle $ states, and the 
hybrids appeared as an extra set of 
$|q\bar qg \rangle +  O(\sqrt{\alpha_s}) (|q\bar q \rangle + |q\bar qg^2 \rangle 
+...)$  
states, which should appear as an ``overpopulation'' of the experimental
meson spectrum relative to the naive $q\bar q$ quark model.

In the bag model the lowest quark mode is a conventional J$^P = 1/2^+$,
but the lowest gluon mode is a (perhaps surprising) J$^P = 1^+$
TE gluon. Combining these lowest lying $q, \bar q$ and $g$ modes, one
finds hybrid basis states with
\begin{displaymath}
{\rm J}^{PC}\bigg|_{bag-model\ hybrids}   =
\bigg( 0^-, 
1^- 
\bigg) \otimes 1^+ = 
1^{--}, 
0^{-+}, 
1^{-+}, 
2^{-+}\ .
\end{displaymath}
Thus the bag model predicts that the lowest lying hybrid multiplet should 
consist of 
these 4 J$^{PC}$, of which the $1^{-+}$ combination is exotic.
Hybrid mass estimates required detailed calculations in which configuration
mixing with the other quark+gluon basis states was included
to $O(\sqrt{\alpha_s})$, and the
resulting truncated Hamiltonian was diagonalized. The results depended
somewhat on the bag model parameters assumed, with masses of
$\approx 1.5$ GeV being typical
\cite{BC,CS}.
Spin dependent splittings ordered the levels as 
$0^{-+} < 1^{-+} < 1^{--} < 2^{-+}$, with a total 
multiplet splitting of 
{\it ca.} 500~MeV
with the usual bag model parameters. Since each of these J$^{PC}$ levels
is a flavor nonet in the $u,d,s$ system, many hybrid states are
predicted that might be experimentally accessible. 

One may also form baryon hybrids, since the basis states $|qqqg\rangle $
contain color singlets. The corresponding bag model calculations of the
spectrum of baryon hybrids 
\cite{BCbary,CHP,GHK}
predict a lowest multiplet of 
$u,d$ ``hybrid baryons'' with a mean mass of about 2 GeV, and a
J$^P$, flavor
content of $(1/2^+ {\rm N})^2$,  $(3/2^+ {\rm N})^2$,  
$(5/2^+ {\rm N})$,  
$(1/2^+ \Delta)$,  
$(3/2^+ \Delta)$. 
The spin-splittings due to quark-gluon and gluon-gluon forces
predict rather large 
overall multiplet splitting of {\it ca.} 500~MeV, resulting in a
$(1/2^+ {\rm N})$ near 1.5 GeV as the lightest hybrid baryon. 
This result led to
the speculation that the N(1440) Roper might be the lightest hybrid baryon.
Of course 
there are no J$^P$ exotics
in the baryons, 
since all J$^P$ can be made
from $qqq$. 
Since $|qqq\rangle \leftrightarrow |qqqg\rangle$ 
configuration mixing is large in the bag model, the
distinction between hybrid and conventional baryons is problematic. One must
simply conclude that, in the context of this model,
there should be an overpopulation of baryons relative
to the simple $|qqq\rangle $ quark model, due to the presence of the extra  
$|qqqg\rangle $ basis states.

\subsection{Flux-tube model hybrids}

In LGT simulations a roughly cylindrical 
region of chaotic
glue fields can be observed between widely separated static color sources.
This ``flux tube'' leads to the confining 
linear potential between color-singlet
$q$ and $\bar q$ that is familiar from quark potential models. 
The ``flux tube model'' \cite{ft_model} is an approximate description
of this state of glue, which is treated as a string of point masses 
``beads'' connected
by a linear potential. This system is treated quantum mechanically, and
has normal modes of excitation which are transverse to the axis between
the (fixed) endpoints of the string. The orbital angular momentum of a
transverse string excitation may be combined with the $q\bar q$ spin and 
orbital angular momentum using rigid body wavefunctions, which leads to
predictions for the quantum numbers of these flux-tube hybrids.
Since the assumption about the nature of excited glue is quite different from
the $1^+$ TE gluon mode of the bag model, one finds a different
spectrum of hybrid states. The lowest flux-tube hybrids are predicted to span
8 J$^{PC}$ levels, all degenerate in the simplest version of the model,
with  
\begin{displaymath}
{\rm J}^{PC}\bigg|_{flux-tube\ hybrids}  =
0^{\pm\mp}, 
1^{\pm\mp}, 
2^{\pm\mp}; 
1^{\pm\pm}\ .
\end{displaymath}
The first 6 of these 
levels have $S_{q\bar q}=1$ and the last 2 have $S_{q\bar q}=0$.

In the earliest mass estimates in the flux tube model
various approximations were made, such as a small oscillation approximation
and an adiabatic quark motion approximation. After several studies of this
system, Isgur, Kokoski and Paton \cite{IKP}
reached their well known estimate of 1.9(1) GeV for 
the mass of this lightest hybrid multiplet.
This work has since been improved upon by 
Barnes, Close and Swanson 
\cite{BCS}
using 
a Hamiltonian Monte Carlo algorithm that does not make the 
small oscillation and adiabatic approximations.
It appears that these approximations gave opposite and 
comparable mass shifts, so their final result was a very similar 1.8-1.9 GeV
for this lightest hybrid multiplet.

Since each of these 8 J$^{PC}$ levels has a flavor nonet of associated states,
the flux tube model predicts a very rich spectrum, with an additional 72 
meson resonances
expected
in the vicinity of 2.0 GeV, 
in addition to the conventional $q\bar q$ quark model states! 

Finally, the very interesting question of the masses and quantum numbers
of hybrid baryons in the flux tube model has only recently been
considered, by Capstick and Page \cite{CapP}.They find that the lightest hybrid 
baryon multiplet
contains degenerate 
$(1/2^+ {\rm N})^2$ and  $(3/2^+ {\rm N})^2$ states
at a mass of 1870(100) MeV, with 
$(1/2^+ \Delta)$,  
$(3/2^+ \Delta)$ and
$(5/2^+ \Delta)$ partners slightly higher in mass.
These conclusions are not so different from the bag model,
which predicted a similar hybrid baryon content, with a 
lightest $(1/2^+ {\rm N})$ hybrid near 1.6 GeV. The most
obvious distinction (other than the 300 MeV difference in the lightest
hybrid baryon's mass) is the high mass $(5/2^+ {\rm N})$ (bag model) 
versus $(5/2^+ \Delta)$ (flux-tube model).

\subsection{LGT and QCD Sum Rules}

These approaches both estimate exotic masses
by evaluating correlation functions of the form
$\langle 0|{\cal O}(\vec x,\tau) {\cal O}^\dagger(0,0)|0 \rangle $, where
$ {\cal O}^\dagger$ is an operator that can couple to the state of interest
from the vacuum. When summed over $\vec x$, at large $\tau$ this quantity
approaches $\kappa \exp( -M_{\cal O}\tau)$, where $M_{\cal O}$ is the mass
of the lightest state created from the vacuum by the operator 
${\cal O}^\dagger$.
Thus by choosing various operators with exotic quantum numbers one may
extract mass estimates for the lightest 
states with those quantum numbers.

Both methods are subject to systematic errors due to approximations. The QCD
sum rules relate these operators to calculable pQCD contributions and to 
VEVs of other operators that are not calculated, but are inferred from 
experiment. Different choices for these parameters, algebra errors 
and uncertainties
in higher-mass contributions have led to a moderately wide 
scatter of results. 
For example, for the $1^{-+}$ exotic, which is of greatest 
phenomenological interest,
the earliest work of Balitsky {\it et al.} in 1982 estimated a mass of
$\approx 1$ GeV. Subsequently
in 1984 
Govaerts {\it et al.} \cite{Lat84} estimated 1.3 GeV, 
Latorre {\it et al.} 
estimated 1.7(1) GeV \cite{Lat84} 
and
2.1 GeV in 1987 \cite{Lat87}. 
The most recent work of
Chetyrkin and Narison \cite{CN} 
finds $\approx 1.6$-$1.7$ GeV, with the radial hybrid only about 
0.2 GeV higher.
This reference also considers
decay couplings; the partial width to $\pi\rho$ is found to be about 300 MeV,
but to $\pi\eta'$ is only about 3 MeV.  As we shall see, this is not what has
been reported
for either experimental exotic $\pi_1$ candidate.

Other exotic quantum numbers have been considered
in QCD sum rules. For example, the $0^{--}$ has been considered 
by several of these references, and is found to have 
a rather
high mass of {\it ca.} 3 GeV. 

Recently LGT groups have presented results for the masses of 
exotic mesons.  The MILC 
collaboration \cite{MILC}
gave results for light 
$1^{-+}$ and $0^{+-}$ exotics, and 
and 
UKQCD
\cite{UKQCD} 
considered these and $2^{+-}$ as well. (These are the three exotics 
predicted to be lightest, and degenerate, in the 
zeroth-order flux tube model.) At present the LGT 
results appear consistent with 
the expectations of the flux-tube model; signals in all 
these channels are observed, with the mass of the $1^{-+}$ 
(the best determined)
being about 2.0(1) GeV. The $0^{+-}$ and $2^{+-}$ may lie somewhat higher,
but this is unclear with present statistics.

The application of 
LGT 
to nonrelativistic heavy quark systems has been the topic of much
recent research, and considerably 
smaller statistical errors follow from the use of a
QCD action derived from a heavy quark expansion. This NRQCD has been applied
to  $1^{-+}$ heavy-quark exotic hybrids, with very interesting results; 
the $1^{-+}$ $b\bar b$-H is predicted to lie at 
$\approx 10.99(1)$ GeV, and the 
$1^{-+}$ $c\bar c$-H charmonium hybrid 
is predicted to lie at $\approx 4.39(1)$ GeV.   
With such small statistical errors in these heavy hybrid 
mass estimates, there is strong motivation for a careful, high statistics 
scan of $R$ near these masses, since models of hybrids anticipate that the
multiplet containing the 
$1^{-+}$ will also possess a $1^{--}$ state nearby in mass.

\section{Hybrid decays}

There appears to be universal agreement that hybrids should exist, and that
the lightest of these states with $u,d$ quarks should include a $1^{-+}$ 
resonance with a mass in the $1.5$-$2$ GeV region, with the higher mass
preferred by LGT and the flux tube model. For the experimental detection
of these states we are faced with the crucial question of what their strong
decay properties are. In the worst case they might be so broad as to
be difficult to identify, a problem familiar from the $f_0$ sector.

Several models of strong decays have been applied to hybrids, and their results
have motivated and directed experimental studies.
The best known is the flux-tube decay model, which was 
applied to exotic hybrids by Isgur, Kokoski and Paton \cite{IKP} and 
subsequently to nonexotic hybrids by Close and 
Page \cite{CP}. 
This model assumes that decays take place by 
$^3$P$_0$ 
$q\bar q$ pair production 
along the length of the flux tube. For 
the {\it unexcited} flux tubes of conventional mesons the predictions are
quite similar to the conventional $^3$P$_0$ model; for hybrids in
the flux tube model this decay assumption allows the calculation
of hybrid meson decay amplitudes.

The orbital angular momentum gives the $q\bar q$ source 
produced during a decay a phase dependence around the original $q\bar q$ 
axis, and the hadronic final states produced are those which have similar
angular dependence.
Naively favorable modes such as $\pi\pi$, $\rho\pi$ and so forth are predicted
to be produced quite weakly due to poor spatial overlap with this 
$\exp(i\phi)$-dependent $q\bar q$ source. The favored modes are those that
have a large L$_z=1$ axial projection, such as an S+P meson pair. This is the
origin of the flux-tube S+P 
selection rule, which in the I=1  $1^{-+}$ case favors
the unusual modes $\pi f_1$ and $\pi b_1$ over 
$\eta\pi$, $\eta'\pi$ and $\rho\pi$, despite their greater
phase space.

In addition to the flux tube decay calculations, there are also QCD
sum rule results (cited above), a decay model that assumes a 
specific relation
between the flux tube excitation and the color vector potential 
\cite{PSS}, and constituent gluon decay amplitude calculations \cite{cgdecay}.
There is general agreement (with some variation between models) 
that in most cases the
flux-tube result of S+P mode dominance in hybrid strong decays is correct.

\begin{table}
\centering
\begin{tabular}{|l|c|c|c|c|c|c|c|c|c|}
\hline
$\pi_1(2000)$
& $\pi f_1$
& $\pi b_1$
&  $\pi \rho$
& $\pi \eta$
&  $\pi\eta'$
&  $a_1\eta $
&  ${\rm K}_1^<{\rm K}$
&  ${\rm K}_1^>{\rm K}$
&  {\rm total}
\\
\hline
\hline
thy.\cite{CP}
&   60
&   170
&  5-20 
&  0-10 
&  0-10
&   50
&   20
& $\sim 125$
& $\approx 450$
\\
\hline
\end{tabular}
\caption{Theoretical two-body partial widths
(MeV) of a $\pi_1(2000)$ flux-tube hybrid.}
\label{table1}
\end{table}

\section{Experimental exotic meson candidates}

\subsection{Introduction}

Since there are only two experimental candidiate exotic meson resonances,
the $\pi_1(1400)$ and the $\pi_1(1600)$,
this section is relatively brief. I will first review the better established
$\pi_1(1600)$, and then discuss the $\pi_1(1400)$. 
Both resonances are reported to have rather 
different properties than theorists expected for exotic hybrid mesons.
Although I will compare these experimental exotic candidiates to theoretical
predictions for exotic {\it hybrids}, they might of course be another
kind of non-$q\bar q$ state or even a misinterpreted nonresonant 
scattering effect.

\subsection{$\pi_1(1600)$}

The best established exotic candidate is the $\pi_1(1600)$. 
Evidence for this state has
been reported in  three channels, 
$b_1 \pi$ (VES \cite{VES_1600}),
$\eta'\pi$ (VES \cite{VES_1600}) and
$\rho\pi$ (VES \cite{VES_1600}, E852 at BNL \cite{E852_1600}). 
The $\rho\pi$ channel is the least controversial, since there are two
independent experiments involved, and clear resonant phase motion against
several well-established $q\bar q$ states is evident. 
The mass and width reported by VES and BNL are consistent,

\begin{equation}
M_{\pi_1} =
\cases{
{\rm 1.61(2)\ GeV}  
&{VES, all three modes}
\cr
1.593\pm 0.008 {+0.029\atop -0.047}\ {\rm GeV}
&{BNL E852, $\rho\pi$},
}
\end{equation}                  

\begin{equation}
\Gamma_{\pi_1} =
\cases{
{\rm 0.29(3)\ GeV} 
&{VES, all three modes}
\cr
0.168 \pm 0.020  {+0.150\atop -0.012}\  {\rm GeV} 
&{BNL E852, $\rho\pi$}.
}
\end{equation}                  

\begin{figure}
\label{fig_1}
\epsfxsize=5truein\epsffile{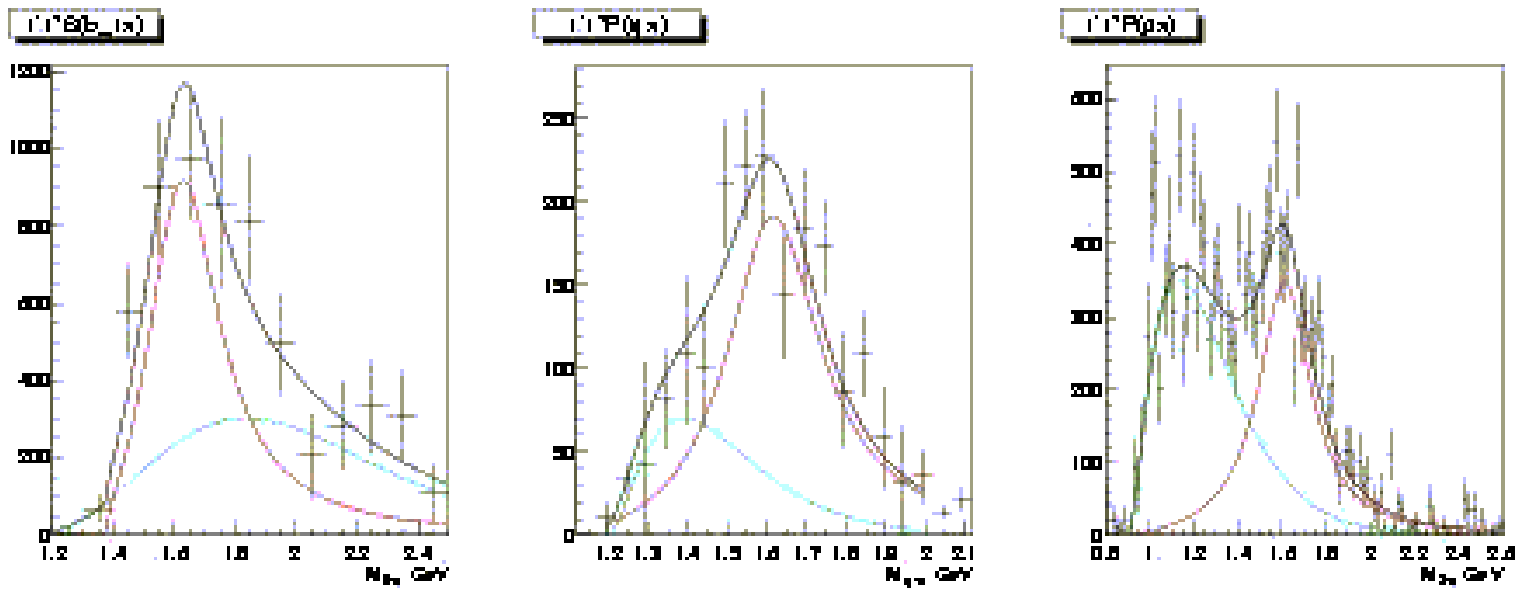}
{\\
Fig.1. VES data \cite{VES_1600} showing the $\pi_1(1600)$ signal in
$b_1\pi$, 
$\eta'\pi$, 
and 
$\rho\pi$.} 
\end{figure}

One difficulty with interpreting this state as a hybrid is the
$\approx 300$-$400$ MeV mass difference between flux-tube and LGT estimates
of $M\approx 1.9$-$2.0$ GeV and the $\pi_1(1600)$ mass.

The evidence for the $\pi_1(1600)$
in VES data in the three channels
$b_1 \pi$,
$\eta'\pi$ and
$\rho\pi$, 
is shown in Fig.2.
The relative branching fractions 
reported by VES 
for these final states 
are

\begin{equation}
\Gamma(\pi_1(1600)\to f ) 
=
\cases{
\equiv 1 
&{$b_1\pi $}
\cr
1.0\pm 0.3
&{$\eta'\pi$}
\cr
1.6\pm 0.4
&{$\rho\pi$} \ .
}
\end{equation}                  

We can see immediately that there is a serious problem here,
as the 
reported relative
branching fractions 
are inconsistent with the predictions of the flux-tube model for 
hybrid decays (Table 1).
The theoretical expectation is that $b_1 \pi$ should be dominant, 
with $\rho\pi$ 
weak and $\eta\pi$ and $\eta'\pi$ very small
\cite{IKP,CP}. Some $\rho\pi$ coupling
is expected in the flux tube model due to different $\rho$ and $\pi$
spatial wavefunctions \cite{CP}, 
but this is expected to be a much smaller effect
in the $\eta\pi$ and $\eta'\pi$ modes. Indeed, there is a generalized G-parity
argument \cite{Page}
that says these would be zero except differences in spatial wavefunctions.
Either these three modes are not all due to a
hybrid exotic, or our 
understanding of hybrid decays is inaccurate. Future experimental studies of
$\rho\pi$ will be especially interesting here, since this channel is easily
accessible for example in photoproduction at the planned HallD facility at
Jefferson Lab \cite{AP,HallD}.

\subsection{$\pi_1(1400)$}

This state is reported in $\eta\pi$, which is a channel with a long and 
complicated history. 
{\it Prima facie} $\eta\pi$
appears to be a very attractive channel in which to search for exotics, 
because there is no spin degree of freedom, and all the 
odd-L $\eta\pi$ channels
are J$^{PC}$-exotic.

The $\eta\pi^o$ channel
was studied by GAMS in 1980 \cite{GAMS} before the idea of J$^{PC}$ exotic
resonances was widely
accepted, and this collaboration was rather surprised to find a significant
(exotic) P-wave. Of course the question was whether this exotic 
partial wave 
showed resonant 
phase motion or was simply a nonresonant background; GAMS had 
insufficient statistics to decide, but speculated that it was probably
nonresonant. 
In a subsequent 1988 study of $\pi^- p \to \eta\pi^o n$ \cite{GAMS88} 
they concluded that $\eta\pi^o$ did indeed support
a 1.4 GeV $1^{-+}$ 
exotic P-wave resonance, although their analysis does not agree with
subsequent studies of the same
$\pi^- p \to \eta\pi^o n$ reaction.
This was followed by a KEK experiment \cite{KEK} 
that concluded that $\eta\pi^-$
did show evidence for a resonant P-wave, {\it albeit} with a 
mass and width consistent with the $a_2(1320)$. Since the 
D-wave $a_2(1320)$ dominated this reaction, there were concerns that
the reported exotic P-wave was actually due to feedthrough of the
large $a_2(1320)$ signal in the partial wave analysis.
This is now believed to be the case, 
perhaps due to the
angular asymmetry of the detector. A subsequent VES experiment also found
a resonant signal in the exotic $\eta\pi^-$ P-wave
\cite{VES_etapi}, but at a rather higher mass.
Their $\pi_1(1400)$ was confirmed in 1997 by BNL experiment E852
\cite{E852_etapi}. Finally, the
Crystal Barrel Collaboration \cite{CB_etapi}
also find that their $\eta\pi\pi$ Dalitz plots
in both charged 
($\eta\pi^o\pi^-$) 
and neutral 
($\eta\pi^o\pi^o$) final states show evidence for  
a broad resonant P-wave exotic,
and fits give a $\pi_1(1400)$ mass and width 
consistent with VES and BNL. The BNL \cite{E852_etapi} and 
Crystal Barrel \cite{CB_etapi} masses and widths are

\begin{equation}
M_{\pi_1} =
\cases{
1.370\pm 0.016 {+0.050\atop -0.030}\ {\rm GeV}
&{BNL E852, $\eta\pi^-$}
\cr
1.400\pm 0.02\pm 0.020 \ {\rm GeV}
&{C.Bar, neutral and charged $\eta\pi$}
\cr
1.360\pm 0.025 \ {\rm GeV}
&{C.Bar, neutral $\eta\pi$},
}
\end{equation}                  

\begin{equation}
\Gamma_{\pi_1} =
\cases{
0.385\pm 0.040 {+0.065\atop -0.105}\ {\rm GeV}
&{BNL E852, $\eta\pi^-$}
\cr
0.310\pm 0.050 {+0.050\atop -0.030} \ {\rm GeV}
&{C.Bar, neutral and charged $\eta\pi$}
\cr
0.220\pm 0.090 \ {\rm GeV}
&{C.Bar, neutral $\eta\pi$}.
}
\end{equation}

\begin{figure}[ht]
\label{fig_2}
\epsfxsize=4truein\epsffile{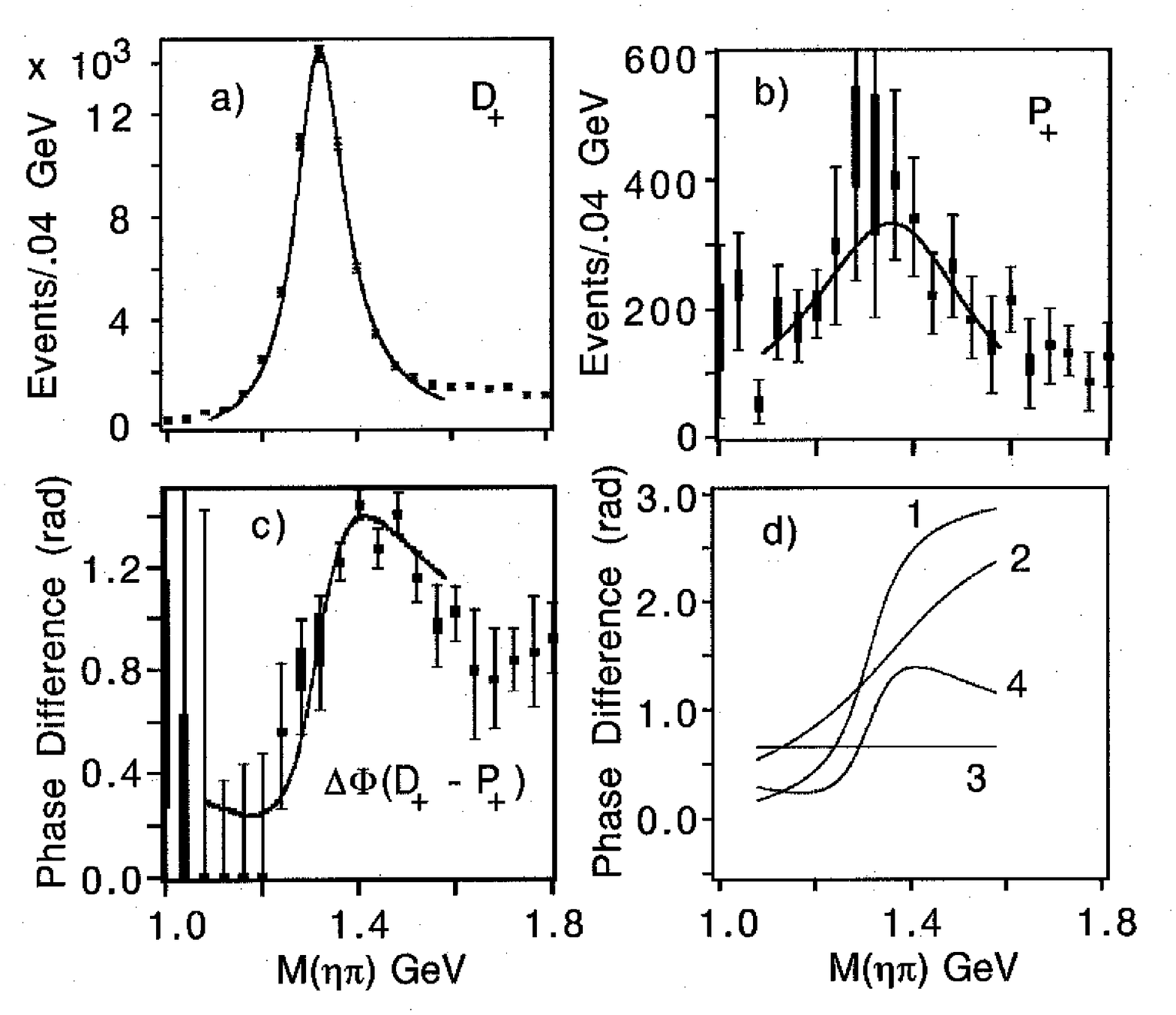}
{\\
Fig.2. The evidence from BNL E852 \cite{E852_etapi} 
for a $1^{-+}$ $\pi_1(1400)$ 
exotic in the $\eta\pi^-$ P-wave. The P-wave modulus is in the upper right, and
the crucial phase motion is shown on the lower left. 
}
\end{figure}

\begin{figure}[ht]
\label{fig_3}
\epsfxsize=4truein\epsffile{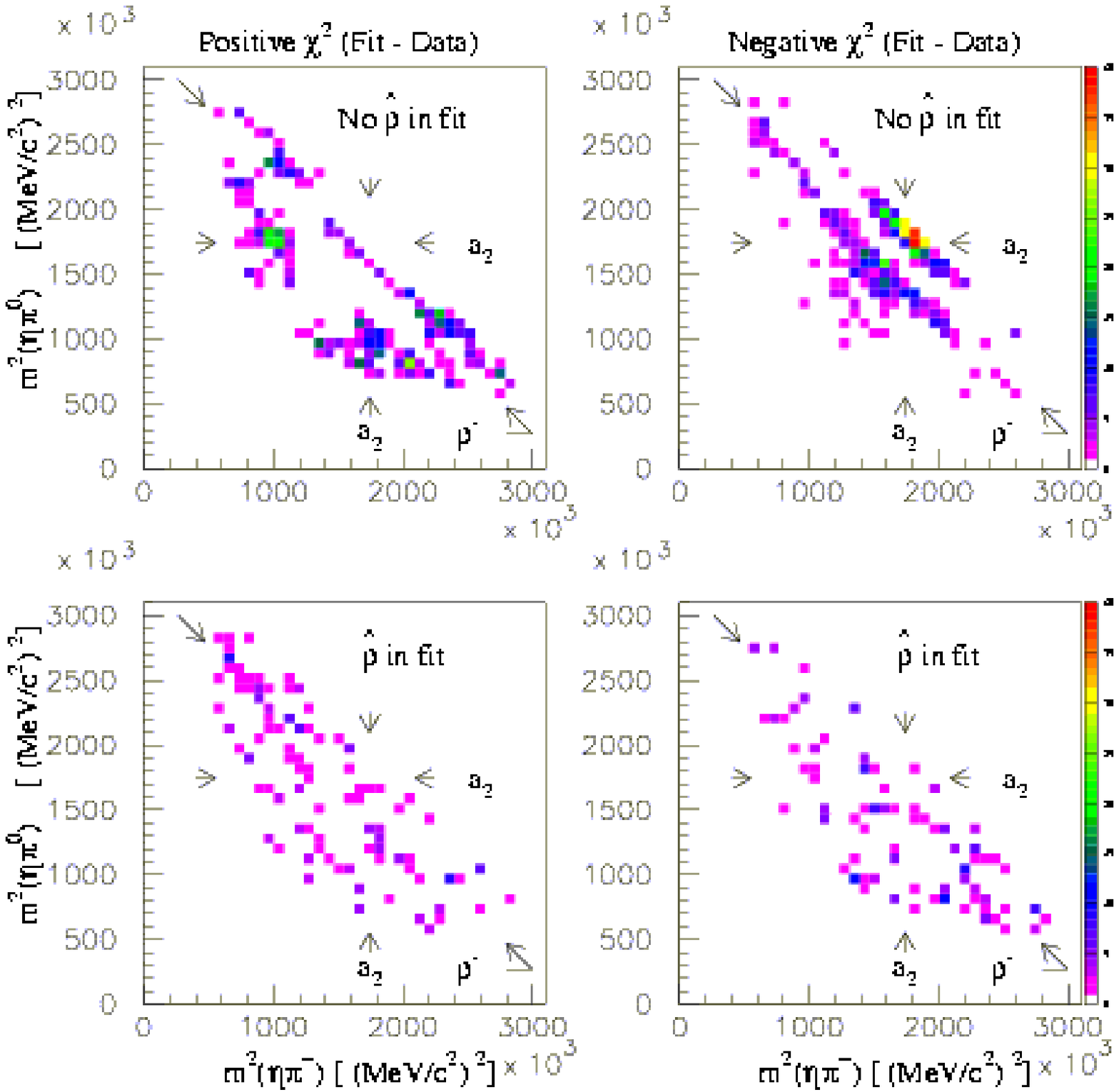}
{\\
Fig.3. Crystal Barrel data \cite{CB_etapi} showing the 
improvement in their description of the $\eta\pi^o\pi^-$ Dalitz
plot when a
$\pi_1(1400)$ is included.
}
\end{figure}

There has been much concern expressed regarding 
various possible experimental 
problems with this rather light and broad $\pi_1(1400)$, 
for example the size 
and energy dependence of backgrounds, and possible nonresonant 
inelastic scattering mechanisms that might mimic a resonance 
\cite{DP,BBS}. One should certainly 
be extremely careful
in establishing the lightest exotic meson.
Nonetheless it is
evident that VES, BNL and Crystal Barrel
have all found evidence for a $\pi_1(1400)$ with comparable mass and width
in $\eta\pi$,
despite theoretical expectations that a light, broad exotic hybrid should not
exist.

\section{Acknowledgements}

It is a great pleasure to thank the organisers for their kind invitation 
to discuss exotics at Meson 2000.
This research was supported in part by the DOE Division of Nuclear Physics,
at ORNL,
managed by UT-Battelle, LLC, for the US Department of Energy
under Contract No. DE-AC05-00OR22725, and by 
the Deutsche Forschungsgemeinschaft DFG
at the University of Bonn and the Forschungszentrum J\"ulich
under contract Bo
56/153-1.


\newpage

\begin{thebibliography}{99}

\bibitem{GN} 
For a very extensive recent review of meson spectroscopy, including 
both conventional and exotic mesons, see
S.Godfrey and J.Napolitano,
Rev. Mod. Phys. 71, 1411 (1999), hep-ph/9811410.

\bibitem{Jaffe1}
R.L.Jaffe and K.Johnson, Phys. Lett. B60, 201 (1976);
R.Jaffe, Phys. Rev. D15, 267 (1977).

\bibitem{WI}
J.Weinstein and N.Isgur,
Phys. Rev. Lett. 48, 659 (1982);
Phys. Rev. D27, 588 (1983);
Phys. Rev. D41, 2236 (1990).

\bibitem{Jaffe2}
R.Jaffe, Phys. Rev. Lett. 38, 195 (1977); {\it err.} 38, 617 (1977).

\bibitem{Richard}
J.-M.Richard, Nucl. Phys. Proc. Suppl. 86, 361 (2000); 
nucl-th/9909030.

\bibitem{BC}
T.Barnes, PhD thesis, Caltech (1977); 
Nucl. Phys. B158, 171 (1979);
T.Barnes and F.E.Close, Phys. Lett. 116B, 365 (1982);
T.Barnes, F.E.Close and F.deViron, Nucl. Phys. B224, 241 (1983).

\bibitem{CS}
M.Chanowitz and S.Sharpe, Nucl. Phys. B222, 211 (1983);
{\it err.} B228, 588 (1983).

\bibitem{BCbary}
T.Barnes and F.E.Close, 
Phys. Lett. 123B, 89 (1983); Phys. Lett. 128B, 277 (1983).

\bibitem{CHP}
C.E.Carlson, T.H.Hansson and C.Peterson,
Phys. Rev. D27, 1556 (1983). 

\bibitem{GHK}
E.Golowich, E.Haqq and G.Karl,
Phys. Rev. D28, 160 (1983); {\it err.} D33, 859 (1986).

\bibitem{ft_model}
N.Isgur and J.Paton, Phys. Rev. D31, 2910 (1985).

\bibitem{IKP}
N.Isgur, R.Kokoski and J.Paton, Phys. Rev. Lett. 54, 869 (1985).

\bibitem{BCS}
T.Barnes, F.E.Close and E.S.Swanson, Phys. Rev. D52, 5242 (1995);

\bibitem{CapP}
S.Capstick and P.R.Page, 
Phys. Rev. D60, 111501 (1999), nucl-th/9904041.

\bibitem{Balitsky}
I.I.Balitsky, D.Diakanov and A.V.Yung, 
Phys. Lett. B112, 71 (1982); a subsequent paper,
Z. Phys. C33, 265 (1986), estimated a rather higher mass.

\bibitem{Gov84}
J. Govaerts, F.deViron, D.Gusbin and J.Weyers, 
Nucl. Phys. B248, 1 (1984).

\bibitem{Lat84}
J.I Latorre, S.Narison, P.Pascual and R.Tarrach, 
Phys. Lett. B147, 169 (1984).

\bibitem{Lat87}
J.I.Latorre, P.Pascual and S.Narison, 
Z. Phys. C34, 347 (1987).

\bibitem{CN}
K.Chetyrkin and S.Narison,
Phys. Lett. B485, 145 (2000), 
hep-ph/0003151v2.

\bibitem{MILC}
C.Bernard {\it et al.} (MILC Collaboration),
Phys. Rev. D56, 7039 (1997).

\bibitem{UKQCD}
P.Lacock {\it et al.} (UKQCD Collaboration),
Phys. Lett. B401, 308 (1997).              

\bibitem{Toussant}
D.Toussant, hep-lat/9909088v2.

\bibitem{CP}
F.E.Close and P.R.Page,
Nucl. Phys. B443, 233 (1995).

\bibitem{PSS}
P.R.Page, E.S.Swanson and A.P.Szczepaniak,
Phys. Rev. D59, 034016 (1999).

\bibitem{cgdecay} 
M.Tanimoto, Phys. Lett. B116, 198 (1982);
A.LeYaouanc {\it et al.}, 
Z. Phys. C28, 309 (1985);
F.Iddir {\it et al.},  
Phys. Lett. B207, 325 (1988);
B433, 125 (1998), hep-ph/9803470;
Yu.S.Kalashnikova,
Z. Phys. C62, 323 (1994).

\bibitem{VES_1600}
V.Dorofeev (VES Collaboration),
in Proceedings of WHS99, hep-ex/9905002.

\bibitem{E852_1600}
G.S.Adams {\it et al.}
(E852 Collaboration),
Phys. Rev. Lett. 81, 5760 (1998).

\bibitem{Page}
P.R.Page, Phys. Lett. B402, 103 (1997).

\bibitem{AP}
A.Afanasev and P.R.Page,
Phys. Rev. D57, 6771 (1998).

\bibitem{HallD}
A.Szczepaniak, these proceedings.

\bibitem{GAMS}
W.D.Apel {\it et al.}, Nucl. Phys. B193, 269 (1981).

\bibitem{GAMS88} 
D.Alde {\it et al.}, Phys. Lett. B205, 397 (1988).

\bibitem{KEK}
H.Aoyagi  {\it et al.}, Phys. Lett. B314, 246 (1993).

\bibitem{VES_etapi}
G.M.Beladidze et al., Phys. Lett. B313,
276 (1993); 
D.V.Amelin {\it et al.}, Phys. Lett. B356, 595 (1995).

\bibitem{E852_etapi}
D.R.Thompson {\it et al.}
(E852 Collaboration),
Phys. Rev. Lett. 79, 1630 (1997);
S.U.Chung {\it et al.}
(E852 Collaboration),
Phys. Rev. D60, 092001 (1999).

\bibitem{CB_etapi}
A.Abele {\it et al.}
(Crystal Barrel Collaboration);
for $\eta\pi^o\pi^-$ see
Phys. Lett. B423, 175 (1998),
for $\eta\pi^o\pi^o$ see
Phys. Lett. B446, 349 (1999).
A review is given by 
C.Amsler, Rev. Mod. Phys. 70,
1293 (1998). The figure is 
the $\eta\pi^o\pi^-$ Dalitz plot,
from 
the CMU Crystal Barrel web site,
http://www.phys.cmu.edu/cb/plots/dalitz.html.


\bibitem{DP}
A.Donnachie and P.R.Page,
Phys. Rev. D58, 114012 (1998),
hep-ph/9807433.

\bibitem{BBS}
T.Barnes, N.Black and E.S.Swanson, 
nucl-th/0007025.
                        

\end{thebibliography}
\end{document}